\documentclass{elsarticle}
\usepackage{graphicx}
\usepackage{amssymb}
\usepackage{amsthm}

\newcommand{\beq}{\begin{eqnarray}}
\newcommand{\eeq}{\end{eqnarray}}

\begin{document}

\begin{frontmatter}

\title{Exact sum rules for inhomogeneous strings}
\author{Paolo Amore}
\ead{paolo.amore@gmail.com}
\address{Facultad de Ciencias, CUICBAS, Universidad de Colima,\\
Bernal D\'{i}az del Castillo 340, Colima, Colima, Mexico}

\begin{abstract}
We derive explicit expressions for the sum rules of the eigenvalues of inhomogeneous strings
with arbitrary density and with different boundary conditions. We show that the sum rule
of order $N$ may be obtained in terms of a diagrammatic expansion, with $(N-1)!/2$ independent
diagrams. These sum rules are used to derive upper and lower bounds to  the energy of the 
fundamental mode of an inhomogeneous string; we also show that it is possible to improve these 
approximations taking into account the asymptotic behaviour of the spectrum and applying the Shanks
transformation to the sequence of approximations obtained to the different orders. 
We discuss three applications of these results.
\end{abstract}

\begin{keyword}
{Helmholtz equation; inhomogeneous string; perturbation theory; collocation method}
\end{keyword}

\end{frontmatter}

\section{Introduction}
\label{sec:intro}

In this paper we consider the problem of obtaining exact results for the spectral zeta functions
of inhomogeneous strings at positive integer values:
\beq
Z(s) = \sum_{n=1} \frac{1}{E_n^s} \ \ \  , \ \ \ s=1,2,\dots
\eeq
where $E_n$ are the eigenvalues of the string, obeying different boundary conditions at the extremities.
Remarkably, the exact calculation of $Z(s)$ for integer values of $s$ does not require the explicit knowledge
of the spectrum, as we have recently pointed out in Ref.\cite{Amore12}. However, the main focus of Ref.\cite{Amore12} was
on the possibility of obtaining a perturbative expansion for $Z(s)$ for $s > d/2$ ($d$ is the number
of dimensions) and use it to evaluate the Casimir energy of the system via its analytic continuation. Here we
restrict our analysis to the one dimensional problem, aiming at obtaining an explicit expression
for the sum rule of arbitrary integer order $n$. The extension of these results to higher dimensions is
discussed in a companion paper~\cite{Amore13b}.

As we have pointed out in \cite{Amore12} there is a good number of examples in the literature where sum rules 
have been obtained for different problems~\cite{Voros80,Steiner85,Berry86,Itzykson86,Steiner87,Steiner87b,Elizalde93,Crandall96,Kvitsinsky96,Voros99,
Mezincescu00,Bender00,Dittmar02,Dittmar05,Arendt09,Dittmar11}: in particular Berry \cite{Berry86} and Crandall \cite{Crandall96}
have used  the sum rules for the modes of certain two dimensional billiards and of one dimensional quantum problems respectively to obtain approximations to the fundamental mode of these systems.  
For the case of a quantum bouncer, Crandall has obtained the energy of the fundamental mode with a
fractional error of $10^{-9}$, using the sum rule of order $30$. 

In this paper we show that it is possible to obtain an explicit expression for the sum rule of a string with arbitrary 
density, and provide a simple set of diagrammatic rules which allow one to write down the expression. The actual calculation
can be conveniently carried out with the help of a computer and therefore sum rules of high order can be 
efficiently calculated. We show that very accurate approximations for the energy of the fundamental modes
can be obtained using a sequence of sum rules. We discuss the application of these results in three examples.

The paper is organized as follows: in Section \ref{sec:sumrules} we derive the general expression for the
sum rule of order $n$ and state a set of diagrammatic rules for $Z(n)$; in Section \ref{sec:appli} we discuss 
the application of the general results of Section \ref{sec:sumrules} to three non--trivial problems; finally,
in Section \ref{sec:concl} we state the conclusions.

\section{Exact sum rules}
\label{sec:sumrules}

In this section we derive exact sum rules for inhomogeneous strings subject to different boundary conditions. 
We consider strings of length $a$ ($|x| \leq a/2$) and  density $\Sigma(x)$ ($\Sigma(x)>0$).

The Helmholtz equation for the inhomogeneous string is
\beq
- \frac{d^2}{dx^2} \psi_n(x)  = E_n \Sigma(x) \psi_n(x) \ ,
\label{Helmholtz1}
\eeq
where $E_n$ and $\psi_n(x)$ are the eigenvalues and eigenfunctions of the string. 
At the extremities of the string, $x=\pm a/2$, the eigenfunctions $\psi_n(x)$ fulfill Dirichlet-Dirichlet, 
Neumann-Neumann, Dirichlet-Neumann, Neumann-Dirichlet or periodic-periodic boundary conditions.

Defining $\phi_n(x) \equiv \sqrt{\Sigma(x)} \ \psi_n(x)$, as discussed in Ref.~\cite{Amore12}, eq.(\ref{Helmholtz1}) 
becomes
\beq
\frac{1}{\sqrt{\Sigma(x)}} \left(- \frac{d^2}{dx^2}\right) \frac{1}{\sqrt{\Sigma(x)}} \phi_n(x) = E_n \phi_n(x)  \ .
\label{Helmholtz2}
\eeq

Therefore $E_n$ and $\phi_n(x)$ are the eigenvalues and eigenfunctions of the hermitian operator 
$\hat{O}\equiv \frac{1}{\sqrt{\Sigma(x)}} \left(- \frac{d^2}{dx^2}\right) \frac{1}{\sqrt{\Sigma(x)}}$; 
the inverse operator is 
$\hat{O}^{-1} \equiv \sqrt{\Sigma(x)} \left(- \frac{d^2}{dx^2}\right)^{-1} \sqrt{\Sigma(x)}$.

Alternatively we may define $\xi_n(x) \equiv \left(- \frac{d^2}{dx^2}\right)^{1/2} \frac{1}{\sqrt{\Sigma(x)}} \phi_n(x)$, which fulfills the equation
\beq
\left(- \frac{d^2}{dx^2}\right)^{1/2}  \frac{1}{\Sigma(x)} \left(- \frac{d^2}{dx^2}\right)^{1/2} \xi_n(x) = E_n \xi_n(x)  \ .
\label{Helmholtz3}
\eeq

In this case $E_n$ and $\xi_n(x)$ are respectively the eigenvalues and eigenfunctions of the hermitian operator 
$\hat{Q}  \equiv \left(- \frac{d^2}{dx^2}\right)^{1/2}  \frac{1}{\Sigma(x)} \left(- \frac{d^2}{dx^2}\right)^{1/2}$; 
its inverse operator is $\hat{Q}^{-1}  \equiv \left(- \frac{d^2}{dx^2}\right)^{-1/2}  \Sigma(x) \left(- \frac{d^2}{dx^2}\right)$.

Observe that
\begin{itemize}
\item the operators $\hat{O}$ and $\hat{Q}$ are isospectral and solving any
of the eqs.(\ref{Helmholtz1}),(\ref{Helmholtz2}) or (\ref{Helmholtz3}), is equivalent to solving the 
remaining  two equations;
\item the eigenvalues of $\hat{O}$ and $\hat{Q}$ grow quadratically in $n$ for $n\rightarrow \infty$, $E_n \propto n^2$;
\item  the spectrum of the inverse operators is bounded both from above and from below, 
$0 < \frac{1}{E_n} \leq \frac{1}{E_1}$, for $n=1,\dots, \infty$;
\item the trace of a hermitian operator is invariant under unitary transformations and therefore 
\beq
Z(s) \equiv Tr \left[\hat{O}^{-1}\right]^s = \sum_{n} \frac{1}{E_n^s}
\eeq
for $s=1,2,\dots$; because of this invariance the trace may be evaluated in any basis, in particular
in the basis of an homogeneous string, $| n \rangle$:
\beq
Z(s) = \sum_n \langle n | \left[\hat{O}^{-1}\right]^s| n \rangle
\label{zetas1}
\eeq
\item because of the asymptotic behavior of $E_n$ for $n \rightarrow \infty$, the trace above is finite
for $s=1,2,\dots$;
\end{itemize}

For practical purposes it is convenient to express Eq.(\ref{zetas1}) directly in terms of the Green's functions of the negative 1D Laplacian, subject to different boundary conditions. 

The one dimensional Green's functions have the general form
\beq
G(x,y) = G_-(x,y) \theta(y-x) + G_+(x,y) \theta(x-y) \ ,
\eeq
where
\beq
G_+(x,y) = G_-(-x,-y) \ \ \ , \ \ \ G_+(y,x) = G_-(x,y)  \nonumber \ .
\eeq

In \ref{appA} we report the explicit expressions for the Green's functions for 
different boundary conditions.

Using these Green's functions we are able to write the spectral zeta function at $s=1,2,\dots$ as:
\beq
Z(n) &=& \int_{-a/2}^{a/2} dx_1 \int_{-a/2}^{a/2} dx_{2}  \dots \int_{-a/2}^{a/2} dx_{n-1} 
\int_{-a/2}^{a_2} dx_n \   \nonumber \\
&\cdot&  G(x_1,x_2) G(x_2,x_3) \dots G(x_{n-1},x_n) G(x_n,x_1) \ \Sigma(x_1) \dots \Sigma(x_n) \ .
\label{zetas2}
\eeq

An equivalent expression for $Z(n)$ is easily obtained using the "x-ordered" product of the
Green's functions:
\beq
Z(n) &=& \int_{-a/2}^{a/2} dx_1 \int_{-a/2}^{x_1} dx_{2}  \dots \int_{-a/2}^{x_{n-2}} dx_{n-1} 
\int_{-a/2}^{x_{n-1}} dx_n \   \nonumber \\
&\cdot& \mathcal{G}(x_1,\dots, x_n) \ \Sigma(x_1) \dots \Sigma(x_n) \ ,
\label{zetas3}
\eeq
where we have defined the $n$-point Green's function
\beq 
\mathcal{G}(x_1,\dots, x_n) \equiv \left[ G(x_1,x_2) G(x_2,x_3) \dots G(x_{n-1},x_n) G(x_n,x_1)\right]_{\mathcal{P}} 
\eeq
and
\beq
\left[ f(x_1,x_2, \dots,x_n) \right]_{\mathcal{P}} \equiv \sum_{permutations}
f(x_{p_1}, \dots , x_{p_n}) \nonumber \ .
\eeq

For example:
\beq
\left[ f(x_1,x_2)\right]_{\mathcal{P}} &\equiv&  f(x_1,x_2) + f(x_2,x_1) \nonumber \\
\left[ f(x_1,x_2,x_3)\right]_{\mathcal{P}} &\equiv&  f(x_1,x_2,x_3) + f(x_1,x_3,x_2) +
f(x_2,x_1,x_3) \nonumber \\
&+& f(x_2,x_3,x_1)+f(x_3,x_1,x_2) + f(x_3,x_2,x_1)
\nonumber  \ .
\eeq

We write explicitly $\mathcal{G}$ up to order $5$:
\beq
\mathcal{G}(x_1) &=& G_+(x_1,x_1) \nonumber \\
\mathcal{G}(x_1,x_2) &=& 2 \ \left[ G_+(x_1,x_2) \right]^2  \nonumber \\
\mathcal{G}(x_1,x_2,x_3) &=& 6 \ G_+(x_1,x_2) G_+(x_1,x_3) G_+(x_2,x_3)  \nonumber  \\
\mathcal{G}(x_1,x_2,x_3,x_4) &=& 8 \ \left[ G_+(x_1,x_2) G_+(x_1,x_4) G_+(x_2,x_3) G_+(x_3,x_4) 
\nonumber \right. \\
&+& \left. G_+(x_1,x_3) G_+(x_1,x_4) G_+(x_2,x_3) G_+(x_2,x_4) \nonumber \right. \\
&+& \left. G_+(x_1,x_2) G_+(x_1,x_3) G_+(x_2,x_4) G_+(x_3,x_4)  \right] \nonumber  \\
\mathcal{G}(x_1,x_2,x_3,x_4,x_5) &=& 10 \ \left[ G_+(x_1,x_4) G_+(x_1,x_5) G_+(x_2,x_3) G_+(x_2,x_5) G_+(x_3,x_4)
\nonumber \right. \\
&+& \left.  G_+(x_1,x_3) G_+(x_1,x_5) G_+(x_2,x_4) G_+(x_2,x_5) G_+(x_3,x_4) \nonumber \right. \\
&+& \left.  G_+(x_1,x_2) G_+(x_1,x_5) G_+(x_2,x_4) G_+(x_3,x_5) G_+(x_3,x_4) \nonumber \right. \\
&+& \left.  G_+(x_1,x_2) G_+(x_1,x_4) G_+(x_2,x_5) G_+(x_3,x_5) G_+(x_3,x_4) \nonumber \right. \\
&+& \left.  G_+(x_1,x_2) G_+(x_1,x_5) G_+(x_2,x_3) G_+(x_4,x_5) G_+(x_3,x_4) \nonumber \right. \\
&+& \left.  G_+(x_1,x_2) G_+(x_1,x_3) G_+(x_2,x_5) G_+(x_4,x_5) G_+(x_3,x_4) \nonumber \right. \\
&+& \left.  G_+(x_1,x_4) G_+(x_1,x_5) G_+(x_2,x_3) G_+(x_2,x_4) G_+(x_3,x_5) \nonumber \right. \\
&+& \left.  G_+(x_1,x_3) G_+(x_1,x_4) G_+(x_2,x_4) G_+(x_2,x_5) G_+(x_3,x_5) \nonumber \right. \\
&+& \left.  G_+(x_1,x_3) G_+(x_1,x_5) G_+(x_2,x_3) G_+(x_2,x_4) G_+(x_4,x_5) \nonumber \right. \\
&+& \left.  G_+(x_1,x_3) G_+(x_1,x_4) G_+(x_2,x_3) G_+(x_2,x_5) G_+(x_4,x_5) \nonumber \right. \\
&+& \left.  G_+(x_1,x_2) G_+(x_1,x_4) G_+(x_2,x_3) G_+(x_3,x_5) G_+(x_4,x_5) \nonumber \right. \\
&+& \left.  G_+(x_1,x_2) G_+(x_1,x_3) G_+(x_2,x_4) G_+(x_3,x_5) G_+(x_4,x_5) \right] \nonumber  \ .
\eeq

We can therefore calculate $Z(n)$ with the diagrammatic rules:

\begin{itemize}
\item Draw n points $x_1, \dots, x_n$ on a line; 
\item Connect each point to any two other points in all possible inequivalent ways excluding
the disconnected diagrams and the diagrams corresponding to a cyclic permutation of the points;  
\item Associate a density $\Sigma(x_i)$ at each point $x_i$ ($i=1,\dots, n$);
\item Associate a factor $G_+(x_i,x_j)$ to each line connecting $x_i$ to $x_j$ ($i<j$);
\item Multiply the result by a factor $2n$, corresponding to the $n$ cyclic permutations
of each inequivalent configuration and to the 2 possible directions in which each diagram can
be traveled;
\item Integrate the expression obtained from the steps above over the internal
points: 

\beq
\int_{-a/2}^{a/2} dx_1 \int_{-a/2}^{x_1} dx_{2}  \dots \int_{-a/2}^{x_{n-2}} dx_{n-1} 
\int_{-a/2}^{x_{n-1}} dx_n \nonumber
\eeq
\end{itemize}

\begin{figure}
\begin{center}
\bigskip\bigskip\bigskip
\includegraphics[width=11cm]{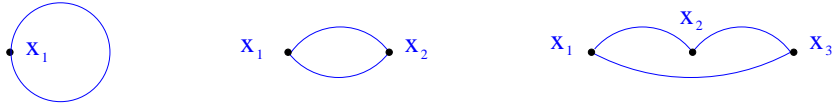}
\caption{Diagrams for $\mathcal{G}(x_1)$, $\mathcal{G}(x_1,x_2)$ and $\mathcal{G}(x_1,x_2,x_3)$.}
\label{Fig_1}
\end{center}
\end{figure}

\begin{figure}
\begin{center}
\bigskip\bigskip\bigskip
\includegraphics[width=13cm]{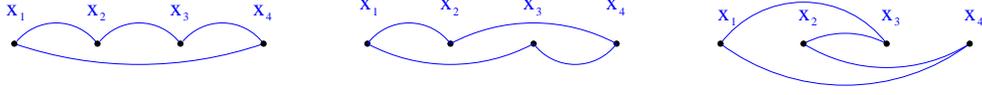}
\caption{Diagrams for $\mathcal{G}(x_1,x_2,x_3,x_4)$.}
\label{Fig_2}
\end{center}
\end{figure}

It is easy to convince oneself that working to order $n$ there are $n!/2 n = (n-1)!/2$ independent diagrams (for $n>2$). For example in Fig.\ref{Fig_1} we plot the diagrams for $\mathcal{G}(x_1)$, $\mathcal{G}(x_1,x_2)$ and $\mathcal{G}(x_1,x_2,x_3)$; in Fig.\ref{Fig_2} we plot the diagrams for $\mathcal{G}(x_1,x_2,x_3,x_4)$: in this case there are $4!/8 = 3$ inequivalent diagrams.

We can now use any of the Eqs.(\ref{zetas1}), (\ref{zetas2}) or (\ref{zetas3}) to evaluate the sum
rules for the string, although Eq.(\ref{zetas3}) is to be preferred for practical purposes, because of its simple diagrammatic representation~\footnote{In Ref.\cite{Amore12} we have used Eq.(\ref{zetas1}) to evaluate explicitly the sum rules of certain strings.}.

It is particularly simple to evaluate $Z(1)$, which reads
\beq
Z^{(DD)}(1) &=& \int_{-a/2}^{+a/2} \left(\frac{a}{4} - \frac{x^2}{a}\right) \Sigma(x) dx \\
Z^{(NN)}(1) &=& \int_{-a/2}^{+a/2} \left( \frac{a}{12} + \frac{x^2}{a} \right) \Sigma(x) dx  \\
Z^{(DN)}(1) &=& \int_{-a/2}^{+a/2} \left( \frac{a}{2} + x \right) \Sigma(x) dx  \\
Z^{(ND)}(1) &=& \int_{-a/2}^{+a/2} \left( \frac{a}{2} - x \right) \Sigma(x) dx  \\
Z^{(PP)}(1) &=& \int_{-a/2}^{+a/2}  \frac{a}{12} \ \Sigma(x) dx  \ .
\eeq

Therefore we obtain the interesting relations:
\beq
Z^{(DD+NN)}(1) &\equiv& Z^{(DD)}(1) + Z^{(NN)}(1) = \frac{a^2}{3} \langle \Sigma \rangle 
\label{sumddnn}
\\
Z^{(DN+ND)}(1) &\equiv& Z^{(DN)}(1) + Z^{(ND)}(1) = a^2 \langle \Sigma \rangle 
\label{sumdnnd}
\\
Z^{(PP)}(1) &=&  \frac{a^2}{12} \langle \Sigma \rangle \ ,
\label{sumpp}
\eeq
where $\langle \Sigma \rangle \equiv \frac{1}{a} \int_{-a/2}^{+a/2} \Sigma(x) dx $ is the average 
density of the string.  Apparently these simple relations have not been found previously.

The diagrammatic rules that we have enunciated allow one to obtain the
spectral zeta functions of a string at large integer values using a computer; 
as we shall see soon, these sum rules are useful
since they allow one to derive approximations to the energy of the ground state. 

As early as 1776, Waring in his "Meditationes Analyticae" considered
the roots $\alpha$, $\beta$, $\gamma$, $\dots$ of an equation (with $\alpha > \beta > \gamma >
\dots$) and discussed the convergence of $(\alpha^{2n} + \beta^{2n} + \gamma^{2n}+\dots)^{1/2n}$
towards the largest root $\alpha$\footnote{This is originally reported by Whittaker and Robinson 
in Ref.~\cite{Whittaker} and by Watson in Ref.~\cite{Watson}. This result appears at pag. 375 
in the edition of 1785, freely available at google books~\cite{Waring1785}.}.
This result may be expressed as~\cite{Berry86}
\beq
E_1 = \lim_{s \rightarrow \infty } Z(s)^{-1/s} \ .
\label{Waring}
\eeq

Little later, in 1781, Euler used the inequalities
\beq
Z(s)^{-1/s} \leq E_1 \leq \frac{Z(s)}{Z(s+1)} \ ,
\label{Euler}
\eeq
to approximate the lowest Bessel zero (see also \cite{Whittaker, Watson, Berry86}). 

Berry \cite{Berry86} and Crandall \cite{Crandall96} have used these formulas to
obtain precise approximations to the energy of the ground state of certain systems: in
particular, Berry in Ref.~\cite{Berry86} has introduced a "semiclassical zeta approximation",
which amounts to evaluate the energy of the ground state with the formula
\beq
E_1 = \left[ Z(2)  - Z^+_1(2)\right]^{-1/2}  \ ,
\label{semiclass1}
\eeq
where, following Berry's notation,
\beq
Z(2) = \sum_{j=1}^N \frac{1}{E_j^2} + \sum_{j=N+1}^\infty \frac{1}{E_j^2}
\equiv Z^{-}_N(2) + Z^{+}_N(2) \ .
\eeq 
Here $Z^{+}_1(2)$ is estimated using a semiclassical approximation for the 
higher part of spectrum. In this way Berry has obtained the lowest energy of 
certain two dimensional billiards within a $1\%$ error.

The approach followed by Berry is particularly convenient in our case, both because
we have at our disposal explicit formulas for the spectral zeta functions at
integer positive values, and because we know the first few terms of the asymptotic behavior of the 
spectrum of the string~\footnote{The asymptotic behavior of the spectrum of an inhomogeneous string
with Dirichlet boundary conditions is $E_n \approx A_1 n^2 + A_2 + A_3/n^2 + \dots$. 
The coefficients $A_1$ and $A_2$ may be obtained using the WKB method; in Ref.\cite{Amore11} 
we have discussed a method which allows one to obtain $A_3$ and higher order coefficients.}.

We may generalize Eq.(\ref{semiclass1}) to
\beq
E_1 = \left[ Z(q)  - Z^+_1(q)\right]^{-1/q}  \ ,
\label{semiclass2}
\eeq
where $q=1,2, \dots$; notice that since $Z(q) > Z_1^+(q)$, one recovers Eq.~(\ref{Waring})
in the limit $q \rightarrow \infty$, neglecting  $Z_1^+(q)$.

The procedure followed by Berry may also be applied to the calculation of the eigenvalues 
of the excited states: in this case the energy of the $n^{th}$ state is obtained as
\beq
E_n = \left[  Z(q-n+1)  - Z^+_{n}(q-n+1) - \sum_{j=1}^{n-1} \frac{1}{E_j^{q-n+1}} \right]^{-1/(q-n+1)} \ .
\eeq

For example, the second eigenvalue is obtained using Eq.(\ref{semiclass2}) and it reads:
\beq
E_2 = \left[  Z(q-1)  - Z^+_2(q-1) -  \left( Z(q)  - Z^+_1(q)\right)^{(q-1)/q}\right]^{-1/(q-1)} \ .
\eeq

Notice that as long as the $Z^+_{n}(q-n+1)$ are evaluated exactly, these equations are also exact;
for sufficiently large $q$ (with $q \gg n$), $Z^+_n(q-n+1)$ may be estimated accurately 
using the asymptotic behavior of the spectrum.

In \ref{appB} we show that, for $n \rightarrow \infty$, for any of the boundary conditions considered
\beq
E_n &\rightarrow& \alpha \epsilon_n + \beta \nonumber \\
&\equiv& \frac{a^2}{\sigma(a/2)^2} \ \epsilon_n + \frac{\int_{-a/2}^{a/2} \frac{4\Sigma(x) \Sigma''(x) - 5 \Sigma'(x)^2}{16 \Sigma(x)^{5/2}} dx }{\int_{-a/2}^{a/2} \sqrt{\Sigma(x)} dx} \ ,
\label{asym}
\eeq
where $\epsilon_n$ are the eigenvalues of a homogeneous string of length $a$ 
and $\sigma(x) \equiv \int_{-a/2}^{x} \sqrt{\Sigma(y)} dy $.  Notice that, for the case of Dirichlet-Dirichlet
boundary conditions, this formula provides the coefficients $A_1$ and $A_2$ of ref.\cite{Amore11}.

Thus we are interested in calculating the semiclassical approximation to $Z_n^+(s)$, which
we call $\tilde{Z}_n^+(s)$:
\beq
\tilde{Z}_n^+(s) &=& \sum_{j=n+1}^\infty \left( \alpha \epsilon_j + \beta \right)^{-s} 
\eeq 

For instance, for $s=1$:
\beq
\tilde{Z}_0^{(DD)+}(1) &=&  \frac{a \coth \left(\frac{a \sqrt{\beta }}{\sqrt{\alpha }}\right)}{2 \sqrt{\alpha } \sqrt{\beta }}-\frac{1}{2 \beta } , \\
\tilde{Z}_0^{(NN)+}(1) &=&\frac{a \tanh \left(\frac{a \sqrt{\beta }}{2 \sqrt{\alpha }}\right)}{4 \sqrt{\alpha } \sqrt{\beta }}+\frac{a \coth\left(\frac{a \sqrt{\beta }}{2 \sqrt{\alpha }}\right)}{4 \sqrt{\alpha } \sqrt{\beta }}-\frac{1}{2 \beta } , \\
\tilde{Z}_0^{(DN)+}(1) &=& \frac{a \tanh \left(\frac{a \sqrt{\beta }}{\sqrt{\alpha }}\right)}{2 \sqrt{\alpha } \sqrt{\beta }} , \\
\tilde{Z}_0^{(PP)+}(1) &=& \frac{a \coth \left(\frac{a \sqrt{\beta }}{2 \sqrt{\alpha }}\right)}{4 \sqrt{\alpha } \sqrt{\beta }}-\frac{1}{2 \beta } . \\
\eeq

Explicit expressions for $s=2,3,\dots$ can be also obtained, although we do not report them here.

When we use the formulas above setting $\beta = 0$ we obtain the interesting relations:
\beq
\tilde{Z}_0^{(DD+NN)+}(1)  &=& \frac{a^2}{3\alpha} = \frac{\sigma(a/2)^2}{3} \\
\tilde{Z}_0^{(DN+ND)+}(1)  &=& \frac{a^2}{\alpha} = \sigma(a/2)^2 \\
\tilde{Z}_0^{(PP)+}(1) &=& \frac{a^2}{12\alpha} = \frac{\sigma(a/2)^2}{12}
\eeq
which have the same form of eqs.(\ref{sumddnn}), (\ref{sumdnnd}), (\ref{sumpp}) apart from 
$\sigma(a/2)^2 \leftrightarrow a^2 \langle \Sigma\rangle$.

We will now discuss a different method to obtain accurate approximations to the 
energy of the fundamental mode of an inhomogeneous string of arbitrary density.
As we have seen, it is possible to obtain explicit expressions for the sum rules 
$Z(n)$ corresponding to different boundary conditions: the only limitation 
to the calculation of $Z(n)$ is the factorial growth of the number of terms in
$\mathcal{G}$ contributing to $Z(n)$ when $n$ is large and the difficulty in performing
analytically the integrations over the coordinates $x_1,\dots, x_n$. 

If we assume that this complications may be overcome up to some $N$, then one 
has a sequence of sum rules, $Z(1),Z(2), \dots, Z(N)$, which can be used to obtain
a sequence of approximations to $E_1$, as explained before.

It is easy to convince oneself that the terms in this sequence converge exponentially
to $E_1$: for example,  we notice that for $s \gg 1$, we have
\beq
Z(s)^{-1/s} \approx E_1 - \frac{E_1}{s}  \left(\frac{E_1}{E_2}\right)^s + \dots
\label{lowerbound}
\eeq
where we have only kept the leading correction. A similar behavior can also 
be inferred for the sequence of ratios $Z(s)/Z(s+1)$:
\beq
\frac{Z(s)}{Z(s+1)} \approx E_1 + E_1 \left(\frac{E_1}{E_2}\right)^s \frac{E_2-E_1}{E_2} + \dots
\label{upperbound}
\eeq
Notice that the lower bound is more accurate than the upper bound.

Sequences with transient behavior of this kind can be efficiently extrapolated using 
the Shanks transformation~\cite{BO78}: in this way, starting with the sequence of 
values of $Z(n)^{-1/n}$ one obtains a new sequence
\beq
\frac{Z(n-1)^{-\frac{1}{n-1}} Z(n+1)^{-\frac{1}{n+1}}-Z(n)^{-\frac{2}{n}}}{Z(n-1)^{-\frac{1}{n-1}} + Z(n+1)^{-\frac{1}{n+1}}-2 Z(n)^{-\frac{1}{n}}} \nonumber
\eeq
which converges more rapidly to $E_1$. The new sequence has $N-2$ terms.

Notice that the Shanks transformation can be applied repeatedly, as long as the sequence at one's 
disposal has at least three terms. Therefore, one can in principle obtain a large gain 
in precision by eliminating several transient behaviors from the original sequence.
The advantage of this procedure is that we are dealing with {\sl exact} sum rules and therefore
we do not have to worry about round-off errors which would necessarily be present 
in a numerical calculation: moreover, the $Z(n)$ are calculated
explicitly as functions of the physical parameters in the problem and therefore the Shanks
transformation will also provide an analytical expression in terms of the physical parameters.

We will see several applications of this method in the following section.

\section{Applications}
\label{sec:appli}

\subsection{Isospectral strings}

Isospectral strings are strings with different densities, but with the same spectrum. A well known example was 
discovered long time ago by Borg ~\cite{Borg46}; this string has a density
\beq
\Sigma(x) = \frac{(1+\alpha)^2}{(1+\alpha (x+1/2))^4} \ \ \ , \ \ \ |x| \leq 1/2 \ ,
\eeq
with $\alpha >-1$. Borg proved that for $\alpha >-1$, all the strings have the same Dirichlet spectrum of a string of constant density, corresponding to $\alpha=0$. 

Using the sum rules obtained before it is easy to see that these strings are only isospectral
to the uniform string for Dirichlet boundary conditions. As a matter of fact already for $s=1$,
the sum rule for Neumann-Neumann, Neumann-Dirichlet, Dirichlet-Neumann and periodic-periodic
all depend on $\alpha$:
\beq
Z^{(NN)}(1) &=& \frac{\alpha  (2 \alpha +3)+3}{18 (\alpha +1)}\\
Z^{(DN)}(1) &=& \frac{\alpha +3}{6 \alpha +6} \\
Z^{(ND)}(1) &=& \frac{1}{6} (2 \alpha +3) \\
Z^{(PP)}(1) &=& \frac{\alpha  (\alpha +3)+3}{36 (\alpha +1)}
\eeq

This result can be better understood noticing that the average density of the string
depends on $\alpha$: $\langle \Sigma \rangle = \frac{\alpha ^2+3 \alpha +3}{3 \alpha +3}$.

Gottlieb has proved in Ref.\cite{Gottlieb02} that, given a string of length $a$ and with density 
$\Sigma(x)$, the strings with density
\beq
\tilde{\Sigma}(x) = \xi'(x)^2 \ \Sigma(\xi(x)) , 
\label{Gottlieb}
\eeq 
with 
\beq
\xi(x) = \frac{a \alpha  (a+2 x)+4 x}{2 a \alpha +4 \alpha  x+4}
\label{moebius}
\eeq
are isospectral to the first string for Dirichlet bc. Notice that $\xi(x)$ maps the interval
$(-a/2,a/2)$ onto itself.
The case discussed by Borg is a special case of eq.(\ref{Gottlieb}) and corresponds 
to $\Sigma(x) = 1$.

It is now easy to check the isospectrality of the strings with Dirichlet bc: it is essential to notice that for the transformation of eq.(\ref{moebius})
\beq
G_+^{(DD)}(x,y) = G_+^{(DD)}(\xi(x),\xi(y)) \ \sqrt{\xi'(x) \xi'(y)}  \ .
\label{green}
\eeq

Using this property we may express the spectral zeta function for the 
string with density eq.(\ref{Gottlieb}) at arbitrary integer values $n$ as
\beq
Z^{(DD)}_{\tilde{\Sigma}}(n) &=& \int_{-a/2}^{a/2} dx_1 \int_{-a/2}^{x_1} dx_{2}  \dots \int_{-a/2}^{x_{n-2}} dx_{n-1} 
\int_{-a/2}^{x_{n-1}} dx_n \   \nonumber \\
&\cdot& \mathcal{G}^{(DD)}(x_1,\dots, x_n) \ \xi'(x_1) \dots \xi'(x_n)  \
\Sigma(\xi(x_1)) \dots \Sigma(\xi(x_n)) \nonumber \\
&=& \int_{-a/2}^{a/2} dx_1 \int_{-a/2}^{\xi_1} d\xi_{2}  \dots \int_{-a/2}^{\xi_{n-2}} d\xi_{n-1} 
\int_{-a/2}^{\xi_{n-1}} d\xi_n \   \nonumber \\
&\cdot& \mathcal{G}^{(DD)}(\xi_1,\dots, \xi_n) \ \Sigma(\xi_1) \dots \Sigma(\xi_n)  = Z^{(DD)}_{\Sigma}(n) \ ,
\eeq
where we have used the notation $Z^{(DD)}_{\tilde{\Sigma}}(n)$ and $Z^{(DD)}_{\Sigma}(n)$ 
for the spectral zeta functions of the strings with Dirichlet bc and with 
density $\tilde{\Sigma}$ and $\Sigma$, respectively. 

This results holds for arbitrary integer $n$ and arbitrary real $\alpha>-1$ and it is consistent
with the isospectrality of the two strings.

It is straightforward to see that the Green's functions corresponding to different
boundary conditions do not obey the transformation (\ref{green}) and therefore they are not
isospectral. 

Notice that starting from order $2$ the sum rules for Neumann bc contain a non polynomial dependence
on $\alpha$. For example $Z^{(NN)}(2)$ is
\beq
Z^{(NN)}(2) &=& \frac{1}{810 \alpha ^4 (\alpha +1)^2} \left[10 \alpha ^8+12 \alpha ^7+93 \alpha ^6
+1422 \alpha ^5+6021 \alpha ^4 \right. \nonumber \\
&+& \left. 12420 \alpha^3 +14220 \alpha ^2+8640 \alpha +2160\right] \nonumber \\
&-&\frac{2 (\alpha +1) (\alpha +2) (\alpha  (\alpha +2)+2) }{3 \alpha ^5} \log(\alpha +1) \nonumber
\eeq
We do not report here higher order sum rules because of their lengthy expressions.

\begin{figure}
\begin{center}
\bigskip\bigskip\bigskip
\includegraphics[width=8cm]{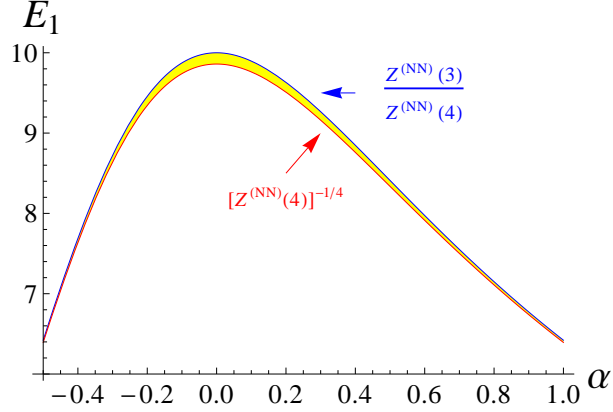}
\caption{Bounds for the energy of the fundamental mode of the Borg string with NN boundary conditions as function of
$\alpha$.}
\label{Fig_borg}
\end{center}
\end{figure}

In Fig~\ref{Fig_borg} we plot the bounds for the energy of the fundamental mode
of the Borg string with Neumann boundary conditions as function of the parameter $\alpha$. The
shaded area is the allowed region. The bounds are obtained using the sum rules of order $3$ and $4$.

\subsection{An exactly solvable string}

We consider a string of density
\beq
\Sigma(x) = \frac{9}{12 x+10} \ \ \ , \ \ \ |x| \leq 1/2 \ ,
\label{Horgan}
\eeq
which belongs to a family of inhomogeneous strings first studied by Horgan and Chan\cite{Horgan99}.
The frequencies of these strings can be calculated with arbitrary precision, since they are
solutions to a trascendental equation. 

In Ref.~\cite{Amore11} we have calculated the first $10000$
Dirichlet eigenvalues of the string (\ref{Horgan}), each with a precision of $200$ digits. 
Using these numerical results, we have extracted the leading asymptotic behavior of 
the Dirichlet spectrum  of this string
\beq
E_n \approx \pi^2 n^2 + \frac{3}{8} - \frac{165}{512 \pi^2 n^2} + \frac{73179}{81920 \pi^4 n^4} - 
\frac{81997443}{14680064 \pi^6 n^6} + \dots 
\label{asymptotic}
\eeq

The first three coefficients of this expansion were also obtained analytically using a WKB-perturbation
expansion, developed in Ref.~\cite{Amore11}.

We can now use these numerical and analytical results to test our sum rules; using Eq.~(\ref{zetas3})
with the help of Mathematica we obtain the exact sum rules for Dirichlet boundary conditions:
\beq
Z^{(DD)}(1) &=& \frac{5}{8}-\frac{2}{3}\ \log (2) \nonumber \\
Z^{(DD)}(2) &=& -\frac{13}{64} + \frac{4}{9} \ \log^2(2)  \nonumber \\
Z^{(DD)}(3) &=& -\frac{105}{1024} -\frac{8}{27} \ \log^3(2) +\frac{7}{24} \  \log(2) \nonumber \\
Z^{(DD)}(4) &=& \frac{131}{46080} +\frac{16}{81}\  \log^4(2) - \frac{7}{27} \  \log^2(2)+ \frac{95}{864} \  \log (2) \nonumber \\
Z^{(DD)}(5) &=& \frac{9521}{589824}-\frac{32}{243} \  \log^5(2)+\frac{35}{162} \  \log^3(2) \nonumber \\
&-& \frac{475}{5184} \ \log^2(2) -\frac{917}{27648} \  \log(2) \nonumber \\
Z^{(DD)}(6) &=& \frac{11466667}{2752512000}+\frac{64}{729} \  \log^6(2) - \frac{14}{81} \  \log^4(2)
+\frac{95}{1296} \  \log^3(2) \nonumber \\
&+& \frac{1897}{34560} \ \log^2(2)-\frac{13183}{368640}  \log (2) \nonumber \\
Z^{(DD)}(7) &=& -\frac{38464127}{31708938240} - \frac{128}{2187} \ \log^7(2) +
\frac{98}{729} \ \log^5(2) - \frac{665}{11664} \  \log^4(2) \nonumber \\
&-& \frac{6713}{103680} \  \log^3(2)+\frac{463043}{9953280} \  \log^2(2)
-\frac{728683}{147456000} \  \log (2) \nonumber \\
Z^{(DD)}(8) &=& -\frac{448469829001}{466121392128000}+\frac{256}{6561} \  \log^8(2)
-\frac{224}{2187} \ \log^6(2) \nonumber \\
&+& \frac{95}{2187} \  \log^5(2) + \frac{3857}{58320} \  \log^4(2) - 
\frac{92749}{1866240}\ \log^3(2) \nonumber \\
&+& \frac{8508391}{5225472000} \  \log^2(2) + \frac{8136221}{1486356480} \  \log(2)\nonumber  \\
Z^{(DD)}(9) &=& -\frac{5652867433}{60881161420800}-\frac{512}{19683} \  \log^9(2)
+\frac{56}{729} \  \log^7(2) - \frac{95}{2916} \  \log^6(2) \nonumber \\
&-& \frac{4837}{77760} \  \log^5(2) + \frac{13261}{276480} \  \log^4(2) +
\frac{7288553}{2322432000} \  \log^3(2) \nonumber \\
&-& \frac{65171959}{5945425920} \  \log^2(2) +\frac{1880324004961}{699182088192000} \  \log (2) \nonumber \ .
\eeq

The factor $\log(2)$ in these expressions is related to the average density of the string,  
$\langle\Sigma \rangle = \int_{-1/2}^{1/2} \Sigma(x) dx = \frac{3}{2} \log(2)$.
Notice that the largest sum rule obtained here corresponds to a diagrammatic 
expansion involving $20160$ inequivalent diagrams.

In Table \ref{table1} we report the error $Z^{(DD)}(q) - Z_{\rm num}^{(DD)}(q)$, where
\beq
Z_{\rm num}^{(DD)}(q) \equiv \sum_{k=1}^{10^4} \frac{1}{{E_k^{\rm (num)}}^q} + 
\sum_{k=10^4+1}^\infty \frac{1}{{E_k^{\rm (asym)}}^q}
\eeq
and $E_k^{\rm (num)}$ are the numerical Dirichlet eigenvalues previously calculated in Ref.~\cite{Amore11}
and $E_k^{\rm (asym)}$ are given by Eq.(\ref{asymptotic}). 

\begin{table}[tbp]
\caption{$Z^{(DD)}(q) - Z_{\rm num}^{(DD)}(q)$. }
\bigskip
\label{table1}
\begin{center}
\begin{tabular}{|c|c|}
\hline
$q$ & $Z^{(DD)}(q) - Z_{\rm num}^{(DD)}(q)$ \\
\hline
$1$ & $6.8 \cdot  10^{-50}$ \\ 
$2$ & $1.2 \cdot  10^{-58}$ \\ 
$3$ & $1.5 \cdot  10^{-67}$ \\ 
$4$ & $1.8 \cdot  10^{-76}$ \\ 
$5$ & $2.1 \cdot  10^{-85}$ \\ 
$6$ & $2.3 \cdot  10^{-94}$ \\ 
$7$ & $2.5 \cdot  10^{-103}$ \\ 
$8$ & $2.6 \cdot  10^{-112}$ \\
$9$ & $2.8 \cdot  10^{-121}$ \\
\hline
\end{tabular}
\end{center}
\bigskip\bigskip
\end{table}

In Table \ref{table2} we report the estimates of $E_1^{(DD)}$ using Eqs.(\ref{Waring}) and (\ref{semiclass2}), where
$Z_{1}^{+(DD)}(q)$ is approximated with $\tilde{Z}_{1}^{+(DD)}(q)$. The underlined digits are exact.

\begin{table}[tbp]
\caption{Estimates of $E_1^{(DD)}$ using Eq.(\ref{Waring}) and Eq.(\ref{semiclass2}). }
\bigskip
\label{table2}
\begin{center}
\begin{tabular}{|c|c|c|}
\hline
$q$ & $\left(Z^{(DD)}(q)\right)^{-1/q}$ & $\left(Z^{(DD)}(q) - \tilde{Z}_{1}^{+(DD)}(q)\right)^{-1/q}$ \\
\hline
1 &  6.13866459 & \underline{10.2}2002206\\
2 &  9.80124983 & \underline{10.218}51148\\
3 & \underline{10}.15503866 & \underline{10.218}20809\\
4 & \underline{10.2}0660399 & \underline{10.2181}3692\\
5 & \underline{10.21}580556 & \underline{10.21811}931\\
6 & \underline{10.21}762510 & \underline{10.21811}486\\
7 & \underline{10.218}00650 & \underline{10.218113}73\\
8 & \underline{10.218}08942 & \underline{10.218113}44\\
9 & \underline{10.2181}0790 & \underline{10.2181133}7\\
\hline
\end{tabular}
\end{center}
\bigskip\bigskip
\end{table}

In Table \ref{table3} we report the estimates of $E_1^{(DD)}$ using repeated 
Shanks transformations of the sequence in the second column of  Table \ref{table2} 
corresponding to Eq.(\ref{Waring}). 

\begin{table}[tbp]
\caption{Estimates of $E_1^{(DD)}$ using repeated Shanks transformations of the 
sequence in the second column of  Table \ref{table2} corresponding to Eq.(\ref{Waring}). }\bigskip
\label{table3}
\begin{center}
\begin{tabular}{|c|c|c|c|}
\hline
$S_1$ & $S_2$ & $S_3$ & $S_4$\\
\hline
\underline{10}.19286707426 & \underline{10.218}09078335 & \underline{10.21811}465291 &   \underline{10.218113344}08 \\ 
\underline{10.21}540206670 & \underline{10.2181}0761972 & \underline{10.2181133}3956 & - \\
\underline{10.21}780418009 & \underline{10.21811}258058 & \underline{10.218113344}10 & - \\
\underline{10.218}07358764 & \underline{10.218113}23885 & - & - \\
\underline{10.2181}0765046 & \underline{10.2181133}2959 & - & - \\
\underline{10.21811}245123 & - & - & - \\
\underline{10.218113}19374 & - & - & - \\
\hline
\end{tabular}
\end{center}
\bigskip\bigskip
\end{table}

In Table \ref{table4} we report the estimates of $E_1^{(DD)}$ using repeated 
Shanks transformations of the sequence in the third column of  Table \ref{table2} 
corresponding to Eq.(\ref{semiclass2}). The energy of the fundamental mode is 
obtained with 17 digits of precision.

\begin{table}[tbp]
\caption{Estimates of $E_1^{(DD)}$ using repeated Shanks transformations of the 
sequence in the third column of  Table \ref{table2} corresponding to Eq.(\ref{semiclass2}). }
\bigskip
\label{table4}
\begin{center}
\begin{tabular}{|c|c|c|c|}
\hline
$S_1$ & $S_2$ & $S_3$ & $S_4$\\
\hline
\underline{10.2181}318565099322 & \underline{10.2181133}603626791 & \underline{10.21811334466}73210 &  \underline{10.21811334466594}08 \\
\underline{10.21811}51161288641 & \underline{10.21811334}57026718 & \underline{10.2181133446659}633 & - \\
\underline{10.218113}5270306264 & \underline{10.218113344}7356183 & \underline{10.218113344665941}1 & - \\
\underline{10.2181133}642743735 & \underline{10.2181133446}706434 & - & - \\
\underline{10.21811334}68298224 & \underline{10.21811334466}62585 & - & - \\
\underline{10.218113344}9084580 & - & - & - \\
\underline{10.2181133446}933714 & - & - & - \\
\hline
\end{tabular}
\end{center}
\bigskip\bigskip
\end{table}

\subsection{A string with rapidly oscillating density}

We consider a string with density
\beq
\Sigma(x) = 2 + \sin \left(\frac{2\pi (x+1/2)}{\epsilon}\right) \ , 
\label{zuazua}
\eeq
with $\epsilon \rightarrow 0^+$ and $|x| \leq 1/2$. This example was studied by Castro and Zuazua 
in Ref.~\cite{CZ00b}, obtaining the asymptotic behavior of the Dirichlet spectrum for $\epsilon \rightarrow 0^+$ 
using the WKB method. More recently, these results have been reproduced in Ref.~\cite{Amore11}, using an 
alternative approach, developed by the author: in particular, the energy of the fundamental mode, which 
had been calculated by Castro and Zuazua to order $\epsilon^4$ has been obtained in Ref.~\cite{Amore11} to order 
$\epsilon^5$ and reads
\beq
E_1^{(DD)} &\approx& \frac{\pi^2}{2} -\frac{\pi ^2}{64} \epsilon^2 + 
\frac{1}{4} \pi  \sin ^2\left(\frac{\pi }{\epsilon }\right) \epsilon^3
-\frac{15 \pi ^2}{1024} \epsilon^4 \nonumber \\
&+& \frac{\pi  \left(5 \sin \left(\frac{4 \pi }{\epsilon }\right)-
116 \cos \left(\frac{2 \pi }{\epsilon}\right)+116\right)}{1024} \epsilon^5 
+ O\left[\epsilon^6\right] \ .
\label{amorewkb}
\eeq

\begin{figure}
\begin{center}
\bigskip\bigskip\bigskip
\includegraphics[width=8cm]{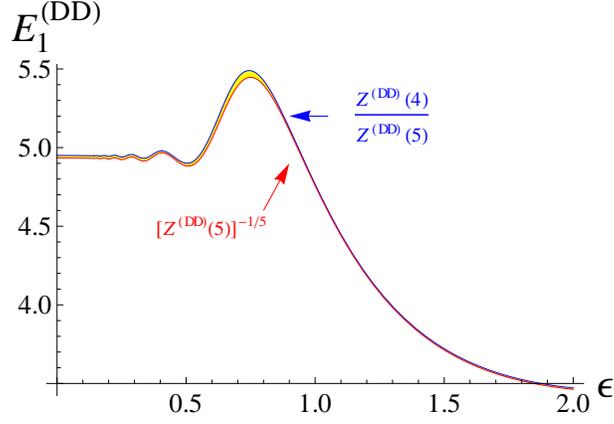}
\caption{Bounds for the energy of the fundamental mode of the 
string (\ref{zuazua}) with Dirichlet boundary conditions as function of
$\epsilon$.}
\label{Fig_zuazuadd}
\end{center}
\end{figure}

\begin{figure}
\begin{center}
\bigskip\bigskip\bigskip
\includegraphics[width=8cm]{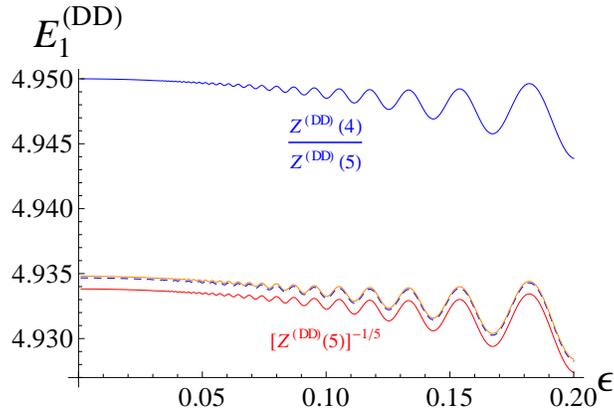}
\caption{Comparison between Eq.(\ref{amorewkb}) (solid line) and 
the Shanks transformation 
$\left[ Z^{(DD)}(3)^{-1/3} Z^{(DD)}(5)^{-1/5} - Z^{(DD)}(4)^{-1/2}\right]/\left[ Z^{(DD)}(3)^{-1/3} + Z^{(DD)}(5)^{-1/5} - 2 Z^{(DD)}(4)^{-1/4}\right]$ (dashed line) as function of $\epsilon$. The lower and upper curves are the
lower and upper bounds respectively.}
\label{Fig_zuazuaddwkb}
\end{center}
\end{figure}

In Fig.\ref{Fig_zuazuadd} we plot the upper and lower bounds obtained with the
sum rules of order 4 and 5 for the 
energy of the fundamental mode of the 
string (\ref{zuazua}) with Dirichlet boundary conditions as function of $\epsilon$. 
In Fig.\ref{Fig_zuazuaddwkb} we compare the exact asymptotic behavior of $E_1^{(DD)}$ for
$\epsilon \ll 1$ of Eq.(\ref{amorewkb}) with the approximation obtained using the Shanks transformation
\beq
S \equiv \frac{Z^{(DD)}(3)^{-1/3} Z^{(DD)}(5)^{-1/5} - Z^{(DD)}(4)^{-1/2}}{ Z^{(DD)}(3)^{-1/3} + Z^{(DD)}(5)^{-1/5} - 2 Z^{(DD)}(4)^{-1/4}} \nonumber \ .
\eeq 

In particular for $\epsilon \rightarrow 0^+$ we have
\beq
S &\approx& 4.9347-0.1543 \epsilon ^2 \nonumber \\
&+&\epsilon ^3 \left(0.3929-0.3929 \cos \left(\frac{2
   \pi }{\epsilon }\right)\right)-0.1463 \epsilon ^4 \nonumber \\
&+& \epsilon ^5 \left(0.0155 \sin \left(\frac{4 \pi }{\epsilon }\right)-0.3605 \cos \left(\frac{2 \pi }{\epsilon}\right)+0.3605\right)
\eeq
which should be compared with the exact asymptotic formula
\beq
E_1^{(DD)} &\approx&  4.9348-0.1542 \epsilon^2 \nonumber \\
&+& \epsilon ^3 \left(0.3927-0.3927 \cos \left(\frac{2 \pi }{\epsilon }\right)\right)-0.1446 \epsilon ^4 \nonumber \\
&+&\epsilon ^5 \left(0.0153 \sin \left(\frac{4 \pi }{\epsilon }\right)-0.3559 \cos \left(\frac{2 \pi }{\epsilon
   }\right)+0.3559\right)+ \dots
\eeq

Notice that $S$ has the correct asymptotic behavior for $\epsilon \rightarrow 0^+$, with coefficients 
which approximate remarkably well the exact coefficients. Notice also that the lower bound 
$\left( Z^{(DD)}(5)\right)^{-1/5}$ is more precise than the upper bound $Z^{(DD)}(4)/Z^{(DD)}(5)$. This
behaviour is consistent with Eqs.(\ref{lowerbound}) and (\ref{upperbound}).

\begin{figure}
\begin{center}
\bigskip\bigskip\bigskip
\includegraphics[width=8cm]{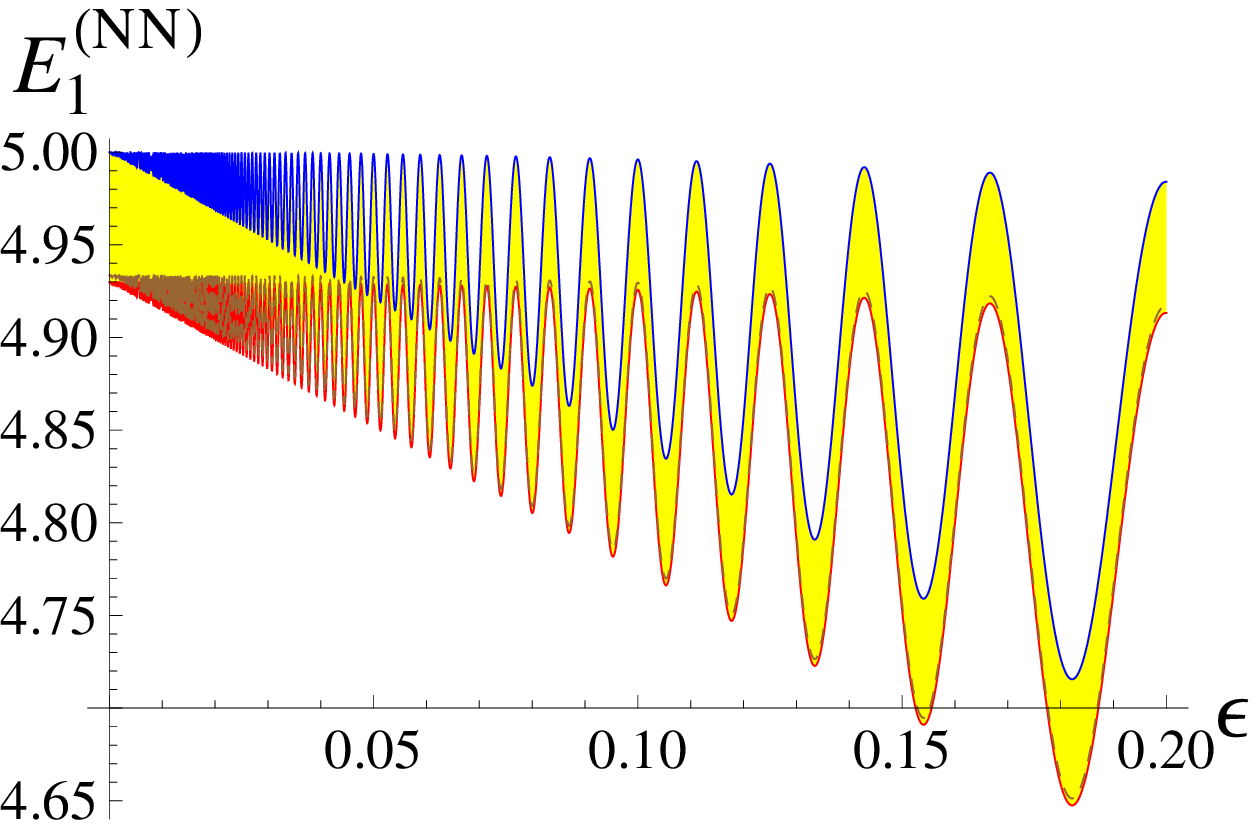}
\caption{Bounds for the energy of the fundamental mode of the 
string (\ref{zuazua}) with Neumann boundary conditions as function of
$\epsilon$.}
\label{Fig_zuazuann}
\end{center}
\end{figure}

In the case of Neumann boundary condition the exact asymptotic behavior for $\epsilon \rightarrow 0^+$ of the fundamental mode of the string (\ref{zuazua})  is not known. It is however straightforward to obtain rigorous 
upper and lower bounds for this energy, using the exact sum rules. In Fig.~\ref{Fig_zuazuann}
we show the bounds obtained using $Z^{(NN)}(3)$ and $Z^{(NN)}(4)$, and the Shanks transformation
obtained using using $Z^{(NN)}(2)$, $Z^{(NN)}(3)$ and $Z^{(NN)}(4)$.
The shaded area corresponds to the allowed region.

If we expand the expression obtained with the Shanks transformation for $\epsilon \rightarrow 0^+$ we
obtain
\beq
E_1^{(NN)} &\approx& 4.9336+\epsilon  \left(0.7852 \cos \left(\frac{2 \pi }{\epsilon}\right)-0.7852\right) 
\nonumber \\
&+& \epsilon^2 \left(-0.3122 \cos \left(\frac{2 \pi }{\epsilon }\right)+0.01562 
\cos \left(\frac{4 \pi}{\epsilon }\right)-0.1084\right) + \dots
\eeq

Therefore we see that for Neumann boundary conditions the fundamental mode of the string
is more sensible to the rapid oscillations of the density: as a matter of fact, $E_1^{(NN)}$ 
contains a term of order $\epsilon$. This term includes an oscillatory contribution 
$\cos \left(\frac{2 \pi }{\epsilon}\right)$; in the case of Dirichlet boundary conditions
the dependence on $\epsilon$ starts at order $\epsilon^2$, while the oscillatory contributions
only start at order $\epsilon^3$. In other words, it is easier to observe the periodicity of the 
density looking at the Neumann rather than Dirichlet spectrum of the string.

\section{Conclusions}
\label{sec:concl}

We have obtained explicit expressions for the sum rules involving the eigenvalues of strings
with arbitrary density for different boundary conditions and we have provided simple diagrammatic
rules which allow one to obtain the expression corresponding to a given order. Despite the factorial
growth of the number of diagrams to a given order, we have derived general expressions up
to order $9$, corresponding to 20160 diagrams, using Mathematica.

These sum rules can be used to obtain precise bounds on the lowest eigenvalue of the string.  
A more accurate determination of this eigenvalue can then be obtained taking into account 
the known asymptotic behaviour of the spectrum and by performing repeated Shanks 
transformations of the sequence of approximations. Since we deal with exact results, no numerical
instability due to round-off errors is ever present.
In this way we have been able to obtain the Dirichlet eigenvalues of a particular string, first discussed by Horgan and Chan, with 17 digits of precision.

For the case of the Borg string, we have proved that the sum rules reduce to the analogous 
sum rules for a homogeneous string only for Dirichlet boundary conditions. 
Therefore the Borg string is isospectral to the homogeneous string only in this case.

For the case of a string with rapidly oscillating density, we have used the sum rules to 
obtain bounds on the lowest eigenvalue: in the case of Dirichlet boundary conditions
we have verified that the sum rule approximate very well  the exact asymptotic behaviour 
of this eigenvalue for arbitrarily rapid oscillations of the density, providing
the exact functional dependence on the physical parameter $\epsilon$.
We have then applied the same method to study the lowest eigenvalue for Neumann bc, for which
the exact asymptotic result is not available, showing that it is more sensible to the 
oscillations of the density.

The extension of these results to higher dimensions is treated in a companion paper~\cite{Amore13b}.

\section*{Acknowledgements}
This research was supported by the Sistema Nacional de Investigadores (M\'exico). 

\appendix
\section{Green's functions}
\label{appA}

In this appendix we derive the explicit expressions for the Green's functions of the
negative laplacian in one dimension and with different boundary conditions.

We need to solve the equation
\beq
-\frac{d^2}{dx^2} G(x,y) = \delta(x-y)
\label{greenf}
\eeq
with $|x| \leq a/2$ and $|y|\leq a/2$.

\subsection{Dirichlet boundary conditions}

The eigenfunctions and eigenvalues of the negative 1D laplacian for Dirichlet boundary 
conditions are
\beq
\psi_n^{(DD)}(x) &=& \sqrt{\frac{2}{a}} \ \sin \frac{n \pi (x+a/2)}{a} \ ,  \\
\epsilon_n^{(DD)} &=& \frac{n^2\pi^2}{a^2} \ .
\eeq

The Green's function is obtained as
\beq
G^{(DD)}(x,y) &=& \sum_{n=1}^\infty \frac{\psi_n^{(DD)}(x)\psi_n^{(DD)}(y)}{\epsilon_n^{(DD)}} 
\eeq
and Eq.(\ref{greenf}) follows from the completeness of the basis $\left\{\psi_n^{(DD)}(x) \right\}$.

It is easy to see that
\beq
G^{(DD)}(x,y) &=& \frac{(a-2 x) (a+2 y)}{4 a} \ \theta (x-y) \nonumber \\
&+& \frac{(a+2 x) (a-2 y)}{4 a}\ \theta(y-x) \nonumber \\
&=& \frac{(a-2 \max [x,y]) (a+2 \min [x,y])}{4 a} 
\eeq

To verify this result we just need to check that, for $x \neq y$, 
$- \frac{d^2}{dx^2} G^{(DD)}(x,y)= 0$ and that 
\beq
\left. -\frac{d}{dx} G^{(DD)}(x,y)\right|_{x\rightarrow y^+} + 
\left. \frac{d}{dx} G^{(DD)}(x,y)\right|_{x\rightarrow y^-} = 1
\eeq
which follows from integrating Eq.(\ref{greenf}) on an arbitrary interval containing $y$.

Notice that
\beq
G^{(DD)}(x,x) &=& \frac{a}{4} - \frac{x^2}{a}
\eeq

\subsection{Neumann boundary conditions}

The eigenfunctions and eigenvalues of the negative laplacian for Neumann boundary conditions are
\beq
\psi_{n,u}^{(NN)}(x) &=& \left\{\begin{array}{ccc}
\sqrt{\frac{1}{a}} & , & n = 0 \ , \ u=1 \\
\sqrt{\frac{2}{a}} \cos \frac{2 n \pi x}{a} & , & n>0 \ , \ u=1 \\
\sqrt{\frac{2}{a}} \sin \frac{(2 n-1) \pi x}{a} & , & n \geq 0 \ , \ u=2 \\
\end{array}
\right. \nonumber \\
\epsilon_{n,u}^{(NN)} &=& \left\{ \begin{array}{ccc}
\frac{4 n^2\pi^2}{a^2} & , & u=1 \\
\frac{(2 n-1)^2\pi^2}{a^2} & , & u=2 \\
\end{array}
\right. 
\eeq

The Green's function is obtained as
\beq
G^{(NN)}(x,y) &=& \sum_{n=0}^\infty  \frac{\psi_{n1}^{(NN)}(x)\psi_{n1}^{(NN)}(y)}{\epsilon_{n1}^{(NN)}} + \sum_{n=1}^\infty  \frac{\psi_{n2}^{(NN)}(x)\psi_{n2}^{(NN)}(y)}{\epsilon_{n2}^{(NN)}}   
\eeq
and Eq.(\ref{greenf}) follows from the completeness of the basis $\left\{ \psi_{nu}^{(NN)}(x)\right\}$. Notice that this expression is formally divergent, because of the 
zero mode which is present in the Neumann spectrum. However the eigenfunction corresponding to 
$n=0$ and $u=1$ is a constant.

The derivation of an explicit expression for $G^{(NN)}(x,y)$ requires a careful discussion
because of the presence of the zero mode. Let us write
\beq
G^{(NN)}(x,y) &=& G_0^{(NN)}(x,y) + \bar{G}^{(NN)}(x,y) 
\eeq
where
\beq
G_0^{(NN)}(x,y) &\equiv& \frac{\psi_{n1}^{(NN)}(x)\psi_{n1}^{(NN)}(y)}{\epsilon_{n1}^{(NN)}} \\
\bar{G}^{(NN)}(x,y) &\equiv& \sum_{n=1}^\infty  \frac{\psi_{n1}^{(NN)}(x)\psi_{n1}^{(NN)}(y)}{\epsilon_{n1}^{(NN)}} + \sum_{n=1}^\infty  \frac{\psi_{n2}^{(NN)}(x)\psi_{n2}^{(NN)}(y)}{\epsilon_{n2}^{(NN)}}   
\eeq
Now
\beq
-\frac{d^2}{dx^2} \bar{G}^{(NN)}(x,y)  &=& \sum_{n=1}^\infty \left[ \psi_{n1}^{(NN)}(x)\psi_{n1}^{(NN)}(y) + \psi_{n2}^{(NN)}(x)\psi_{n2}^{(NN)}(y) \right] \nonumber \\
&=& \delta(x-y) -  \psi_{01}^{(NN)}(x)\psi_{01}^{(NN)}(y)  \nonumber \\
&=& \delta(x-y) -  \frac{1}{a} \ .
\eeq

Therefore, for $x \neq y$ one must have
\beq
-\frac{d^2}{dx^2} \bar{G}^{(NN)}(x,y) &=& -  \frac{1}{a} \ .
\eeq

To ensure that the rhs of Eq.(\ref{greenf}) is obtained we also need to impose that
\beq
\left. -\frac{d}{dx} G^{(NN)}(x,y)\right|_{x\rightarrow y^+} + \left. \frac{d}{dx} G^{(NN)}(x,y)\right|_{x\rightarrow y^-} = 1
\eeq

However since $G_0^{(NN)}(x,y)$ is a constant, the equation above may be cast 
directly as
\beq
\left. -\frac{d}{dx} \bar{G}^{(NN)}(x,y)\right|_{x\rightarrow y^+} + 
\left.  \frac{d}{dx} \bar{G}^{(NN)}(x,y)\right|_{x\rightarrow y^-} = 1
\eeq

It is easy to see that 
\beq
\bar{G}^{(NN)}(x,y) &=& 
\frac{\left(a^2+6 a (y-x)+6 \left(x^2+y^2\right)\right)}{12 a} \ \theta(x-y)\nonumber \\
&+& \frac{\left(a^2+6 a (x-y)+6 \left(x^2+y^2\right)\right)}{12 a}  \ \theta (y-x) \nonumber \\
&=& \frac{\left(a^2 -6 a  |x-y| +6 \left(x^2+y^2\right)\right)}{12 a} 
\eeq
and
\beq
\bar{G}^{(NN)}(x,x) &=& \frac{a}{12} + \frac{x^2}{a}
\eeq

\subsection{Mixed boundary conditions: Dirichlet-Neumann}

The eigenfunctions and eigenvalues of the negative laplacian for Dirichlet-Neumann boundary conditions are
\beq
\psi_{n}^{(DN)}(x) &=& \psi_{2n-1}^{(DD)}\left(\frac{-x+a/2}{2}\right) \ , \ n \geq 1 \nonumber \\
\epsilon_{n}^{(DN)} &=& \frac{(2 n-1)^2\pi^2}{4a^2} 
\eeq

The Green's function is obtained as
\beq
G^{(DN)}(x,y) &=& \sum_{n=1}^\infty \frac{\psi_{n}^{(DN)}(x)\psi_{n}^{(DN)}(y)}{\epsilon_{n}^{(DN)}}  \nonumber 
\eeq

The derivation of the Green's function for this case is completely analogous to what done before for  
Dirichlet boundary conditions and therefore we only report the results:
\beq
G^{(DN)}(x,y)  &=& (x+a/2) \theta(y-x) + (y+a/2)  \theta(x-y) \nonumber \\
&=& \left(\min [x,y] +a/2\right)
\eeq
and
\beq
G^{(DN)}(x,x) &=& x+\frac{a}{2}
\eeq

\subsection{Mixed boundary conditions: Neumann-Dirichlet}

The eigenfunctions and eigenvalues of the negative laplacian for Neumann-Dirichlet boundary conditions are
\beq
\psi_{n}^{(ND)}(x) &=& \psi_{2n-1}^{(DD)}\left(\frac{x+a/2}{2}\right) \ , \ n \geq 1 \nonumber \\
\epsilon_{n}^{(ND)} &=& \frac{(2 n-1)^2\pi^2}{4a^2} 
\eeq

The Green's function is obtained as
\beq
G^{(ND)}(x,y) &=& \sum_{n=1}^\infty \frac{\psi_{n}^{(ND)}(x)\psi_{n}^{(ND)}(y)}{\epsilon_{n}^{(ND)}}  \nonumber 
\eeq

Once again we only report the result:
\beq
G^{(ND)}(x,y) &=& (-x+a/2) \theta(x-y) + (-y+a/2)  \theta(y-x)  \nonumber \\
&=& \left( - \max [x,y] +a/2 \right)
\eeq
and
\beq
G^{(ND)}(x,x) &=& -x+\frac{a}{2}
\eeq

\subsection{Periodic boundary conditions}

The eigenfunctions and eigenvalues of the negative laplacian for periodic boundary conditions are
\beq
\psi_{n}^{(PP)}(x) &=& \left\{\begin{array}{ccc}
\sqrt{\frac{1}{a}} & , & n = 0 \ , \ u=1 \\
\sqrt{\frac{2}{a}} \cos \frac{2 n \pi x}{a} & , & n>0 \ , \ u=1 \\
\sqrt{\frac{2}{a}} \sin \frac{2 n \pi x}{a} & , & n \geq 0 \ , \ u=2 \\
\end{array}
\right. \nonumber \\
\epsilon_{n,u}^{(PP)} &=& \frac{4 n^2\pi^2}{a^2}
\eeq

The Green's function is obtained as
\beq
G^{(PP)}(x,y) &=& \sum_{n=0}^\infty \frac{\psi_{n1}^{(PP)}(x)\psi_{n1}^{(PP)}(y)}{\epsilon_{n1}^{(PP)}}  
+ \sum_{n=1}^\infty \frac{\psi_{n2}^{(PP)}(x)\psi_{n2}^{(PP)}(y)}{\epsilon_{n2}^{(PP)}} \nonumber
\eeq

In this case the same considerations done for the case of Neumann bc apply and we have
\beq
\bar{G}^{(PP)}(x,y)  &=& \frac{a^2+6 a (x-y)+6 (x-y)^2}{12 a} \theta(y-x) \nonumber \\
&+& \frac{a^2+6 a (y-x)+6 (x-y)^2}{12 a} \theta(x-y) \nonumber \\
&=& \frac{a^2 -6 a |x-y| + 6 (x-y)^2}{12 a}
\eeq
and
\beq
G^{(PP)}(x,x) &=& \frac{a}{12}
\eeq

\section{Asymptotic laws}
\label{appB}

We derive here the leading asymptotic behavior of the spectrum of an inhomogeneous string 
subject to any of the boundary conditions discussed in this paper. We follow the  
discussion of Ref.~\cite{Amore11} and consider the operator $\hat{O} = \frac{1}{\sqrt{\Sigma}} 
\left( - \frac{d^2}{dx^2}\right) \frac{1}{\sqrt{\Sigma}}$, whose spectrum coincides with the 
spectrum of a string of density $\Sigma(x)$.

Let $\epsilon_n$ and $\psi_n(x)$ be the eigenvalues and eigenfunctions of the negative 1d laplacian:
\beq
- \frac{d^2  \psi_n}{dx^2} = \epsilon_n \psi_n(x) \nonumber \ .
\eeq

Define:
\beq
\Psi_n(x) \equiv \sqrt{\frac{a}{\sigma(a/2)}} \  \Sigma(x)^{1/4} \ \psi_n\left(a \frac{\sigma(x)}{\sigma(a/2)} - \frac{a}{2}\right)
\eeq
where $\sigma(x) \equiv \int_{-a/2}^{x} \sqrt{\Sigma(y)} dy$. Notice that 
$\Psi_n(x)$ and  $\psi_n(x)$ obey the same boundary conditions only
for the Dirichlet case.

Then
\beq
\hat{O} \Psi_n(x) = \left[ \tilde{\epsilon}_n + \frac{4\Sigma(x) \Sigma''(x) -5 \Sigma'(x)^2}{16\Sigma(x)^3}\right] \Psi_n(x) 
\eeq
where $\tilde{\epsilon}_n = \frac{a^2}{\sigma(a/2)^2} \ \epsilon_n$ are the eigenvalues of a homogeneous string of length $\sigma(a/2)$.

Taking into account that $\tilde{\epsilon}_n \propto n^2$ for $n \rightarrow \infty$, we see that the
$\Psi_n(x)$ tend to become eigenfunctions of $\hat{O}$ in this limit; we can therefore introduce the operator
\beq
\hat{P} \equiv \hat{O} - \frac{4\Sigma(x) \Sigma''(x) -5 \Sigma'(x)^2}{16\Sigma(x)^3} \ ,
\eeq
such that
\beq
\hat{P} \Psi_n(x) = \tilde{\epsilon}_n \Psi_n(x) \ ,
\eeq
and write 
\beq
\hat{O} = \hat{P} +  \frac{4\Sigma(x) \Sigma''(x) -5 \Sigma'(x)^2}{16\Sigma(x)^3} 
\eeq

We can now apply perturbation theory and calculate the corrections to the eigenvalues of the
string, treating the second term in $\hat{O}$ as a perturbation:
\beq
E_n^{(0)} &=& \tilde{\epsilon}_n \\
E_n^{(1)} &=& \langle n | V | n \rangle = \int_{-a/2}^{a/2}  \Psi_n^2(x) \ 
\frac{4\Sigma(x) \Sigma''(x) -5 \Sigma'(x)^2}{16\Sigma(x)^3} \ dx \\
\dots &=&  \dots \nonumber
\eeq

In the limit $n \rightarrow \infty$ 
\beq
E_n^{(1)} \rightarrow \frac{1}{\sigma(a/2)} \int_{-a/2}^{a/2} \frac{4\Sigma(x) \Sigma''(x) -5 \Sigma'(x)^2}{16\Sigma(x)^{5/2}} dx + O\left[ \frac{1}{n^2} \right] \ .
\eeq

We can thus read off the asymptotic coefficients $A_1$ and $A_2$ of the string; the calculation of the 
higher order coefficients can be performed in a similar way. For the reader interested in this calculation we
refer to Ref.\cite{Amore11}, where we have derived the analytical expression for $A_3$ (see Eq.(53)).

Therefore
\beq
E_n \rightarrow \tilde{\epsilon}_n +  \frac{1}{\sigma(a/2)} \int_{-a/2}^{a/2} \frac{4\Sigma(x) \Sigma''(x) -5 \Sigma'(x)^2}{16\Sigma(x)^{5/2}} dx + O\left[ \frac{1}{n^2} \right] + \dots
\eeq

\end{document}